\documentclass[11pt]{article}
\usepackage{amssymb}
\usepackage{hyperref}

\usepackage{graphicx}
\usepackage{amsmath}
\usepackage{upgreek}
\newtheorem{theorem}{Theorem}

\newtheorem{conjecture}[theorem]{Conjecture}

\newcommand {\Z}{\mathbb{Z}}

\newcommand {\g}{\mathfrak{g}}

\newcommand {\Spin}{\mathrm{Spin}}

\renewcommand {\O}{\mathcal{O}}

\newcommand{\Q}{\mathbb{Q}}
\renewcommand{\L}{{\cal L}}
\newcommand{\dbar}{\bar \partial}

\renewcommand{\L}{\mathcal{L}}
\newcommand{\sd}{\partial}
\newcommand{\SO}{\mathrm{SO}}

\renewcommand{\O}{\mathcal{O}}
\newcommand{\C}{\mathbb{C}}

\newcommand{\halpha}{\hat{\alpha}}

\newcommand{\OGr}{\mathrm{OGr}}

\newcommand{\ralpha}{\upalpha}

\newcommand{\setE}{\mathrm{E}}

\newcommand{\K}{\mathcal{K}}
\newcommand{\Maps}{\mathpzc{Maps}}

\newcommand{\bT}{\mathbf{T}}

\DeclareMathAlphabet{\mathpzc}{OT1}{pzc}{m}{it}

\newcommand{\itwo}{\frac{\infty}{2}}

\newcommand{\hA}{A}

\newcommand{\cone}{\mathcal{C}}
\newcommand{\qu}{\mathcal{Q}}
\newcommand{\tspace}{\mathcal{X}}

\newcommand{\Pf}{\mathrm{Pf}}

\newcommand{\Eq}{\mathrm{Eq}}

\begin{document}

\title{A formula for the partition function of the $\beta\gamma$ system on the cone pure spinors} 
\author{M.V. Movshev\\
Mathematics Department, Stony Brook University,\\
 Stony Brook NY, 11794-3651, USA\\
\texttt{mmovshev@math.sunysb.edu}}
\date{\today}
\maketitle

\begin{abstract}
In this note,  we propose a closed  formula for the partition function $Z(t,q)$ of the $\beta\gamma$ system on the cone of pure spinors.  We give the answer  in terms of theta functions, $q$-Pochhammer symbols and Eisenstein series. 
\end{abstract}

\section{Introduction}
The $\beta\gamma$ system on the cone of pure spinors $\cone$ is an integral part of the version of the  string theory invented by N. Berkovits \cite{Berdef}. 
$\cone$ is an eleven-dimensional  subvariety  in  16-dimensional linear space  with coordinates $ \lambda,p_i,w_{ij},\quad1\leq i,j\leq 5, w_{ij}=-w_{ji}$
 defined by the equations 
\begin{equation}\label{E:relexpl0}
\begin{split}
&\lambda p_{i}-\Pf_{i}(w)=0\quad {i}=1,\dots,5,\\
&pw=0.\\
\end{split}
\end{equation}
$\Pf_{i}(w), 1\leq i\leq 5$ are the principal Pfaffians of $w$.
The action is common to all $\beta\gamma$ systems:
\[S(\beta,\gamma)=\int_{\Sigma}\langle\dbar\beta,\gamma\rangle.\]
The field  $\beta$ is a smooth map $\beta:\Sigma\rightarrow \cone$, where $\Sigma$ is a Riemann surface,  $\gamma$ is a smooth section of the pullback $\beta^*T_{\cone}^*\otimes T^*_{\Sigma}$. We advise the reader  to consult \cite{LocMov} for the notation  and discussion of issues related to definition of a $\beta\gamma$ systems on the nonsmooth $\cone$.

  In \cite{LocMov},  a geometric construction of the space of states  $H^{i+\itwo},i=0,\dots,3$ of this system was presented. It is a properly regularized space of the semi-infinite local cohomology of the space of polynomial maps $\Maps(\C^{\times},\cone)$. The support of the local cohomology lies at $\Maps(\C,\cone)$. The space  $\cone$ is an affine cone over $\OGr(5,10)$. $\C^{\times}\times \Spin(10)$ is the groups of symmetries $\cone$, where $\C^{\times}$ acts by dilations. The group $\C^{\times}\times\bT\times \Spin(10)$ acts by the symmetries of the pair  $\Maps(\C,\cone)\subset \Maps(\C^{\times},\cone)$. The factor $\bT\cong\C^{\times}$ corresponds to  loop rotations. The action of $\C^{\times}\times \bT \times \Spin(10)$ survive the regularization and continue to act on $H^{i+\itwo}$. It turns out (see \cite{LocMov}) that the formal character 
 \begin{equation}\label{E:virtchar}
 Z(t,q,z)=\sum_{i=0}^3(-1)^i\chi_{H^{i+\itwo}}(t,q,z)
 \end{equation} is well defined as an element in $\Z((t,z_1,\dots,z_5))((q))\cap \Q(t,z_1,\dots,z_5)((q))$, where $z_1,\dots,z_5$ are the coordinates on the Cartan subgroup $\bT^5\subset \Spin(10)$. More precisely, $Z(t,q,z)$ is a limit of coefficients of an infinite matrix product  \cite{LocMov}. The matrix product simplifies when $z=1$ ( \ref{E:matrixproduct}). We use it to derive the formula for $Z(t,q):=Z(t,q,1)$. 
 
It is necessary to state from the outset that the analysis presented in this note 
is based on experimentations with the formula for $Z_N^{N'}(t,q)$ using {\it Mathematica} for finite $N,N'$ and extrapolation of the found structures to infinite $N$s. Though somewhat loose in justification, the result looks  convincing because it passes a number of consistency checks.
 \paragraph{Acknowledgment}  
The author would like thank N.Berkovits,  A.S. Schwarz, E. Witten  for their  interest and  helpful conversations about this work.

\section{The formula}
As it is mentioned in the abstract,  $Z(t,q)$ will be expressed in therms of some standard special functions. We  start with reviewing their definitions.

\paragraph{The functions used in the formula}

Recall that the $q$-Pochhammer symbol is an infinite product
\[(t;q)_{\infty}:=\prod_{n\geq 0}(1-tq^n).\]
It is  used to write concisely  the three-term identity 
\begin{equation}\label{E:thetadef}
\theta(t, q)=(1-t^{-1})(q,q)_{\infty}(q/t,q)_{\infty}(qt,q)_{\infty}
\end{equation}
for the theta function
\[\theta(q,t):=\sum_{n\in\Z}(-1)^nq^{\frac{n(n+1)}{2}}t^{n}.\]
Another ingredient  of the formula are  theta functions with characteristics:
\begin{equation}\label{E:epsilondef}
\epsilon_k(t, q)=
\sum_{n\in\Z}(-1)^n
q^{\frac{7n(n+1)}{2}+kn}t^{7n+k}
=t^k \left(1-q^{-k}t^{-7}\right)  \left(q^7,q^7\right)_{\infty } \left(q^{7-k}t^{-7},q^7\right)_{\infty } \left(q^{k+7} t^7,q^7\right)_{\infty }, k=0,\dots,6.
\end{equation}
The proposed answer will also depend on the Eisenstein series
\[E_2(q):=1-24\sum_{n=1}^{\infty}\frac{nq^n}{1-q^n},\quad E_4(q):=1+240\sum_{n=1}^{\infty}\frac{n^3q^n}{1-q^n}.
\]
Define  linear combinations of $\epsilon_k(q,t)$:
\[\theta_1(t,q):=-q^3\epsilon_3(q,t)-q^4\epsilon_4(q,t),\]
\[\theta_2(t,q):=q^2\epsilon_2(q,t)+q^5\epsilon_5(q,t) \text{, and}\]
\[\theta_3(t,q):=-q\epsilon_1(q,t)-q^6\epsilon_6(q,t)\]
and introduce  abbreviation $\theta_k(q):=\theta_k(1,q), k=1,2,3$.

The following conjecture contains the promised formula.
\begin{conjecture}
The partition function $Z(t,q)$ has the form
\begin{equation}\label{E:mainelliptic}
Z(t,q)=\frac{a(q)\theta_1(t,q)+b(q)\theta_2(t,q)+c(q)\theta_3(t,q)}{t^6\theta(t,q)^{11}}=:\Psi(t,q)
\end{equation}
where the string functions $a,b,c$ are
\begin{equation}\label{E:a}
\begin{split}
&a:=
\frac{1}{768 (q,q)_{\infty}^4}
(2744 q^2 E_2(q) \theta_3(q) \theta_2''(q)-2744 q^2 E_2(q) \theta_2(q) \theta_3''(q)-343 q E_2(q)^2 \theta_3(q) \theta_2'(q)+3626 q E_2(q) \theta_3(q) \theta_2'(q)\\
&+343 q E_2(q)^2 \theta_2(q) \theta_3'(q)-5194 q E_2(q) \theta_2(q) \theta_3'(q)+98 E_2(q)^2 \theta_2(q) \theta_3(q)-476 E_2(q) \theta_2(q) \theta_3(q)\\
&+42 q E_4(q) \theta_3(q) \theta_2'(q)-42 q E_4(q) \theta_2(q) \theta_3'(q)-12 E_4(q) \theta_2(q) \theta_3(q)-65856 q^3 \theta_2''(q) \theta_3'(q)\\
&+65856 q^3 \theta_2'(q) \theta_3''(q)-29400 q^2 \theta_3(q) \theta_2''(q)+37632 q^2 \theta_2'(q) \theta_3'(q)+10584 q^2 \theta_2(q) \theta_3''(q)\\
&-25725 q \theta_3(q) \theta_2'(q)+18333 q \theta_2(q) \theta_3'(q)+1350 \theta_2(q) \theta_3(q)),
\end{split}
\end{equation}
\begin{equation}\label{E:b}
\begin{split}
&b:=-\frac{1}{768 (q,q)_{\infty}^4}
(2744 q^2 E_2(q) \theta_3(q) \theta_1''(q)-2744 q^2 E_2(q) \theta_1(q) \theta_3''(q)-343 q E_2(q)^2 \theta_3(q) \theta_1'(q)+2842 q E_2(q) \theta_3(q) \theta_1'(q)\\
&+343 q E_2(q)^2 \theta_1(q) \theta_3'(q)-5194 q E_2(q) \theta_1(q) \theta_3'(q)+147 E_2(q)^2 \theta_1(q) \theta_3(q)-546 E_2(q) \theta_1(q) \theta_3(q)\\
&+42 q E_4(q) \theta_3(q) \theta_1'(q)-42 q E_4(q) \theta_1(q) \theta_3'(q)-18 E_4(q) \theta_1(q) \theta_3(q)-65856 q^3 \theta_1''(q) \theta_3'(q)\\
&+65856 q^3 \theta_1'(q) \theta_3''(q)-29400 q^2 \theta_3(q) \theta_1''(q)+56448 q^2 \theta_1'(q) \theta_3'(q)+1176 q^2 \theta_1(q) \theta_3''(q)\\
&-17325 q \theta_3(q) \theta_1'(q)+2205 q \theta_1(q) \theta_3'(q)+225 \theta_1(q) \theta_3(q)),
\end{split}
\end{equation}
\begin{equation}\label{E:c}
\begin{split}
&c:=\frac{1}{768 (q,q)_{\infty}^4}
(2744 q^2 E_2(q) \theta_2(q) \theta_1''(q)-2744 q^2 E_2(q) \theta_1(q) \theta_2''(q)-343 q E_2(q)^2 \theta_2(q) \theta_1'(q)+2842 q E_2(q) \theta_2(q) \theta_1'(q)\\
&+343 q E_2(q)^2 \theta_1(q) \theta_2'(q)-3626 q E_2(q) \theta_1(q) \theta_2'(q)+49 E_2(q)^2 \theta_1(q) \theta_2(q)-70 E_2(q) \theta_1(q) \theta_2(q)\\
&+42 q E_4(q) \theta_2(q) \theta_1'(q)-42 q E_4(q) \theta_1(q) \theta_2'(q)-6 E_4(q) \theta_1(q) \theta_2(q)-65856 q^3 \theta_1''(q) \theta_2'(q)\\
&+65856 q^3 \theta_1'(q) \theta_2''(q)-10584 q^2 \theta_2(q) \theta_1''(q)+18816 q^2 \theta_1'(q) \theta_2'(q)+1176 q^2 \theta_1(q) \theta_2''(q)\\
&-9261 q \theta_2(q) \theta_1'(q)+1533 q \theta_1(q) \theta_2'(q)+27 \theta_1(q) \theta_2(q)).
\end{split}
\end{equation}
\end{conjecture}

\section{Supporting evidences}\label{E:evidences}
\paragraph{The matrix product presentation for $Z(t,q)$}

 It was  established in \cite{LocMov} that $Z(t,q)$ is the limit in the sense of formal power series convergence  of a certain infinite matrix product.
To state the result let us fix some additional notations:
\begin{equation}\label{E:hilbertini}
\begin{split}
&B_0^1:=\frac{1+3t+t^2}{(1-t)^{8}(1-qt)}, \quad  \hA_0^0:=\frac{1+5t+5t^2+t^3}{(1-t)^{11}} ,\\
&K(t,q):=\left(
\begin{array}{cc}
 \frac{t \left(t^2+3 t+1\right)}{(t-1)^7 (q t-1)} & \frac{\left(t^2+3 t+1\right) (t^3+q^2)-5 q (t+1) t^2}{q^2 (t-1)^7 (q t-1)} \\
 \frac{t (t+1) \left(t^2+4 t+1\right)}{(t-1)^{10}} & \frac{\left(t^3+5 t^2+5 t+1\right) (t^3+q^2)-q \left(5 t^2+14 t+5\right) t^2}{q^2 (t-1)^{10}} \\
\end{array}
\right),
\end{split}
\end{equation}

\begin{equation}\label{E:matrixproduct}
\left(
\begin{array}{c}
 B_0^{r+1} \\
 \hA_0^r \\
\end{array}
\right):=
K(q^{r}t,q) \cdots K(qt,q) 
\left(
\begin{array}{c}
 B_0^1 \\
 \hA_0^0 \\
\end{array}
\right),
\end{equation}
\[ \hA_N^{N'}(t,q):=\hA_0^{N'-N}(tq^{N},q).\]
It was  verified in  \cite{LocMov}  that the limit of  
\begin{equation}\label{E:explicitform}
Z_N^{N'}(t,q):=\hA_N^{N'}(t,q)t^{4-4N}q^{-2+4N-2N^2}, N<0
\end{equation}
$N\to -\infty, N'\to \infty$ coincides with $Z(t,q)$ (\ref{E:virtchar}).

\paragraph{Poles  of $Z(t,q)$}

The rational function $Z_N^{N'}(t,q)$ has a fairly complicated structure. Still, experiments with {\it Mathematica} show that $Z_N^{N'}(t,q)$ has poles of multiplicity $\dim_{\C}\cone=11$ precisely  at $q^{-N}, \dots, q^{-N'}$. It is natural to conjecture that in the limit $N\to -\infty, N'\to \infty$ this pattern persists and $Z$ is a meromorphic function for $|q|<1$ with poles at $t=q^n, n \in \Z$ of multiplicity $11$.

\paragraph{$Z(t,q)$ and a line bundle of degree $7$}

 If this conjecture is true, then the product
\begin{equation}\label{E:prodmod}
\Theta(t,q):=t^6Z(t,q)\theta(t,q)^{11}
\end{equation}
is an analytic function for $t\neq 0, |q|<1$.

One of the results of  \cite{LocMov} is that  $Z(t,q,g)$ as a formal power series in $q$ satisfies 
\begin{equation}\label{E:bgequations1}
\frac{Z(qt,q,g)}{Z(t,q,g)}=\frac{t^4}{q^2},
\end{equation}
\begin{equation}\label{E:bgequations2}
 \frac{Z(1/t,q,g^{-1})}{Z(t,q,g)}=-t^8.
\end{equation}
It follows from the functional equations
\begin{equation}\label{E:thetasmall}
\theta(qt,q)=-\theta(t,q)/(qt),\quad \theta(1/t,q)=-t\theta(t,q)
\end{equation}   that $\Theta(t,q)$
obeys
\begin{equation}\label{E:prodequation}
\frac{\Theta(tq,q)}{\Theta(t,q)}=-\frac{1}{q^7t^7},
\end{equation}
\begin{equation}\label{E:prodequation2}
 \frac{\Theta(1/t,q)}{\Theta(t,q)}=t^7
\end{equation}
Stated differently, equation (\ref{E:thetasmall})
 says  that $\theta$   is a holomorphic section of a line bundle $\L$ of degree  one on the elliptic curve $\C^{\times}/\{q^k\}$. 
 To say that $\Theta$ satisfy   (\ref{E:prodequation})   is equivalent to  saying  that  $\Theta$ is a section of  $\L^{\otimes 7}$.
  The  space of global sections of $\L^{\otimes 7}$   has a basis $\epsilon_k, k=0,\dots,6$ (\ref{E:epsilondef}). Symmetry condition  (\ref{E:prodequation2}) determines a subspace in the span of $\epsilon_k$  of dimension  three with  a basis $\theta_i,i=1,2,3$.
As a consequence, we get
\begin{equation}\label{E:thetachar}
\Theta(t,q)=a(q)\theta_1(t,q)+b(q)\theta_2(t,q)+c(q)\theta_3(t,q),
\end{equation}
which is equivalent to (\ref{E:mainelliptic}). 
\paragraph{Equations on coefficients  $a,b,c$}
It remains to determine  $a,b,c$. Denote the right-hand side in (\ref{E:mainelliptic}) by $\Psi(t,q)$. Identity
\begin{equation}\label{E:seriesofequations}
\lim_{t\to 1}\partial^k_tZ(t,q)(1-t)^{11}=
\lim_{t\to 1}\partial^k_t\Psi(t,q)(1-t)^{11}
,\quad  k=k_1,k_2,k_3
\end{equation}
produces  three linear equations for $a(q),b(q),c(q)$. Denote by $x$  the vector $(a,b,c)$. We can write the system of equations on $x$  in the matrix form         
\begin{equation}\label{E:axzeq}
z^t=Ax^t.
\end{equation}  The coefficients of the  matrix $A$  depend only on the functions $\theta_i$ and can be explicitly computed. The vector $z$ is more complicated because it depends on the unknown function $Z$.
But  if we manage to compute three Taylor coefficients of $Z(t,q)$ with respect to the variable $t$,  we can easily find the functions  $a,b,c$.

\paragraph{Coefficients of the matrix $A$}

  Let us find the matrix $A$ first. The convenient choice for $k_i$ in (\ref{E:seriesofequations}) is $0,2,4$. The function $\theta(t,q)$ has a zero of order one at $t=1$ so that $\frac{\theta(t,q)}{t-1}$ is regular.  Denote \[\theta_0(q):=\lim_{t\to1}\frac{\theta(t,q)}{1-t}.\]
The right-hand-side of fist equation $\Eq_{0}$ (\ref{E:seriesofequations}), the $k=0$ case, becomes
\begin{equation}\label{E:eqq1}
\lim_{t\to1}\Psi(t,q)(1-t)^{11}=\frac{a\theta_1+b\theta_2+c\theta_3}{\theta_0^{11}}.
\end{equation}
The second equation $\Eq_{2}$, corresponding to $k=2$ \footnote{$\Eq_{1}$  is proportional  to $\Eq_{0}$}, contain $t$-partial derivatives of functions at $t=1$. Note that  $\theta$ and $\epsilon_k$ satisfy the heat equations:
\[t^2\sd^2_t\theta+ t \sd_t\theta=2q \sd_q\theta,
 \quad   t^2\sd^2_t\epsilon_k+8 t\sd_t\epsilon_k-k(k+7) \epsilon_k=14 q \sd_q\epsilon_k. \]
In addition, linear combinations $\theta_i$ of $\epsilon_k$ satisfy  (\ref{E:prodequation2}). These equations allow to express $t$-derivatives of $\theta_i$ at  $t=1$ in terms of $q$-derivatives of $\theta_i$.  Thus equations  \ref{E:thetadef} and the results of \cite{NikosBagis} imply that 
 \[\theta_0(q)=-(q,q)^3_{\infty},\quad \lim_{t\to1} \partial_t \frac{\theta(t,q)}{1-t}=-\theta_0(q),\quad \lim_{t\to1} \partial^2_t \frac{\theta(t,q)}{1-t}=\theta_0\times\left(\frac{23}{12}+\frac{E_2}{12}\right),\]
 
 \[\lim_{t\to1} \partial^3_t \frac{\theta(t,q)}{1-t}=\theta_0\times\left(-\frac{11}{2} -\frac{1}{2}E_2\right),\] 
 
\[\lim_{t\to1} \partial^4_t \frac{\theta(t,q)}{1-t}=\theta_0\times\left(\frac{1689}{80} + \frac{23}{8}E_2 + \frac{1}{48}E_2^2 - \frac{1}{120}E_4\right),\]

\[\lim_{t\to1} \partial^5_t \frac{\theta(t,q)}{1-t}=\theta_0\times\left(
-\frac{1627}{16}
-\frac{145}{8}E_2
-\frac{5}{16}E_2^2 
+\frac{1}{8}E_4\right).\]
Similarly
\[\theta_1(q):=\theta_1(1,q)=2(q^3,q^7)_{\infty}(q^4,q^7)_{\infty}(q^7,q^7)_{\infty},\quad \partial_t\theta_1(1,q)=-7/2\theta_1(q),\quad \partial^2_t\theta_1(1,q)=16\theta_1(q)+ 14q\partial_q\theta_1(q),\]
\[\partial^3_t\theta_1(1,q)=-90\theta_1-189q\partial_q\theta_1,\quad \partial^4_t\theta_1(1,q)=600\theta_1+2268q\partial_q\theta_1+196q^2\partial^2_q\theta_1,\]

\[\theta_2(q):=\theta_2(1,q)=-2(q^2,q^7)_{\infty}(q^5,q^7)_{\infty}(q^7,q^7)_{\infty},\quad \partial_t\theta_2(1,q)=-7/2\theta_2,\quad \partial^2_t\theta_2(1,q)=18\theta_2+ 14q\partial_q\theta_2,\]
\[\partial^3_t\theta_2(1,q)=-117\theta_2-189q\partial_q\theta_2,\quad \partial^4_t\theta_2(1,q)=900\theta_2+2324q\partial_q\theta_2+196q^2\partial^2_q\theta_2,\]
\[
\theta_3(q):=\theta_3(1,q)=2(q,q^7)_{\infty}(q^6,q^7)_{\infty}(q^7,q^7)_{\infty},\quad \partial_t\theta_3(1,q)=-7/2\theta_3,\quad \partial^2_t\theta_3(1,q)=22\theta_3+ 14q\partial_q\theta_3,\]
\[\partial^3_t\theta_3(1,q)=-171\theta_3-189q\partial_q\theta_3,\quad \partial^4_t\theta_3(1,q)=1524\theta_3+2436q\partial_q\theta_3+196q^2\partial^2_q\theta_3.\]
After simplifications $\Eq_{2}$ and $\Eq_{4}$ become
\begin{equation}\label{E:eqq2}
\begin{split}
& \lim_{t\to1}\sd_t^2 \Psi(t,q)(1-t)^{11}=-\frac{1}{12 \theta_0^{11}}\times\\
&\times((11E_2\theta_1-168q\theta_1'-23\theta_1)a +(11E_2\theta_2-168q\theta_2'-47\theta_2)b+(11E_2\theta_3-168q\theta_3'-95\theta_3)c),
\end{split}
\end{equation}
\begin{equation}\label{E:eqq3}
\begin{split}
& \lim_{t\to1}\sd_t^4 \Psi(t,q)(1-t)^{11}=\frac{1}{240 \theta_0^{11}}\times\\
&\times ((-18480q E_2\theta_1'+605E_2^2\theta_1-990E_2\theta_1+22E_4\theta_1+47040q^2\theta_1''+58800q\theta_1'+363\theta_1)a\\
&(-18480q E_2\theta_2'+605E_2^2\theta_2-3630E_2\theta_2+22E_4\theta_2+47040q^2\theta_2''+72240q\theta_2'+3003\theta_2)b\\
&(-18480q E_2\theta_3'+605E_2^2\theta_3-8910E_2\theta_3+22E_4\theta_3+47040q^2\theta_3''+99120q\theta_3'+14043\theta_3)c).
\end{split}
\end{equation}
The matrix $A$ consists of  coefficients of $a,b,c$ from equations (\ref{E:eqq1},\ref{E:eqq2},\ref{E:eqq3}). 
\paragraph{The vector $z$ components}
Computation of the vector $z$ from (\ref{E:axzeq}) relies on the extrapolation of the results obtained  with {\it Mathematica}. We  use  that $Z(t,q)$ has presentation as a limit of (\ref{E:explicitform}), which makes it possible  computation of  the $q$-series $\lim_{t\to1}\sd_t^i Z(t,q)(1-t)^{11}$ with arbitrary precision. Due to integrality of the coefficients, $\lim_{t\to1}\sd_t^i Z_N^{N'}(t,q)(1-t)^{11}\mod q^r$ stabilizes for sufficiently large $N$ and $N'$.
The first computation shows that 
\[\begin{split}
&\frac{1}{12}\lim_{t\to1}Z(t,q)(1-t)^{11}=\\
&=1 + 22 q + 275 q^2 + 2530 q^3 + 18975 q^4 + 122430 q^5 + 702328 q^6 + 3661900 q^7 + 17627775 q^8+\dots O(q^{20}).
\end{split} \]
The database
 \href{http://oeis.org/A023020}{http://oeis.org}   hints that   these series
is the generating function for  the sequence     \href{http://oeis.org/A023020}{A023020}  . $a_n$ is the  number of partitions of n into parts of 22 kinds. 
The generating function coincides with the Taylor expansion of $\frac{1}{(q,q)^{22}_{\infty}}$.

Another computation with {\it Mathematica} shows that
 \[
 \frac{1}{98}(q,q)_{\infty}^{22} \lim_{t\to1}\sd_t^2 Z(t,q)(1-t)^{11}=
 1/6 + q + 3 q^2 + 4 q^3 + 7 q^4 + 6 q^5 + 12 q^6 + 8 q^7 + 15 q^8+\cdots O(q^{20}).
 \]
 If we drop the term $1/6$, the sequence  $a_n$ of Taylor coefficients of the remaining series is 
  \href{http://oeis.org/A000203}{A000203} . 	$a(n) = \sigma(n)$, the sum of the divisors of $n$. The generating function  for $a_n$ is $(1-E_2(q))/24$.
 The same way we find that
 \[  (q,q)^{22}_{\infty}\lim_{t\to1}\sd_t^4 Z(t,q)(1-t)^{11}=\frac{42}{5}-12E_2(q)+4E_2^2(q)-\frac{2}{5}E_4+O(q^{20}).\]
 To summarize, the vector $z$ in (\ref{E:axzeq}) is conjecturally equal
 \[\left(\frac{12}{(q,q)^{22}_{\infty}},\frac{20-4E_2(q)}{(q,q)^{22}_{\infty}}, \frac{\frac{42}{5}-12E_2(q)+4E_2^2(q)-\frac{2}{5}E_4}{(q,q)^{22}_{\infty}} \right)\]
 The matrix $A$ has the determinant 
 \[\Delta=\frac{8\tilde{\Delta}}{\theta_0 ^{33}}\] 
  \[\begin{split}
& \tilde{\Delta}=(-343 q^3 \theta_3 \theta_1'' \theta_2'+343 q^3 \theta_2 \theta_1'' \theta_3'+343 q^3 \theta_3 \theta_1' \theta_2''\\
&-343 q^3 \theta_2 \theta_1' \theta_3''-343 q^3 \theta_1 \theta_2'' \theta_3'+343 q^3 \theta_1 \theta_2' \theta_3''+98 q^2 \theta_2 \theta_3 \theta_1''\\
&+98 q^2 \theta_3 \theta_1' \theta_2'-294 q^2 \theta_2 \theta_1' \theta_3'-147 q^2 \theta_1 \theta_3 \theta_2''+196 q^2 \theta_1 \theta_2' \theta_3'\\
&+49 q^2 \theta_1 \theta_2 \theta_3''+42 q \theta_2 \theta_3 \theta_1'-126 q \theta_1 \theta_3 \theta_2'+84 q \theta_1 \theta_2 \theta_3'+6 \theta_1 \theta_2 \theta_3).
 \end{split}\]
 
It is not hard to check with  {\it Mathematica}  that
\[ \tilde{\Delta}=-48(q,q)_{\infty}^{15}+O(q^{300}).\]
We conjecture that this is an exact equality. We use it and the   Kramer's rule to derive from (\ref{E:axzeq}) the formulas (\ref{E:a},\ref{E:b},\ref{E:c}).
\section{Some  consistency checks}
Note that by construction  the function $\Psi$ (\ref{E:mainelliptic}) satisfies  equations (\ref{E:bgequations1},\ref{E:bgequations2}). $Z(t,q)$ is the solution of the same set of equations . The $q$-expansion of $\Psi$ is 
\[\begin{split}
&\Psi(t,q)=-\frac{t^3+5 t^2+5 t+1}{(t-1)^{11}}\\
&-\frac{q \left(46 t^3+86 t^2+86 t+46\right)}{(t-1)^{11}}\\
&+\frac{q^2 \left(t^{11}-11 t^{10}+55 t^9-181 t^8-567 t^7-947 t^6-947 t^5-567 t^4-181 t^3+55 t^2-11 t+1\right)}{(t-1)^{11} t^4}\\
&+\frac{2 q^3 \left(8 t^{13}-65 t^{12}+195 t^{11}-143 t^{10}-1011 t^9-2657 t^8-3917 t^7-3917 t^6-2657 t^5-1011 t^4-143 t^3+195 t^2-65 t+8\right)}{(t-1)^{11} t^5}\\
&-\frac{q^4 \left(-126 t^{15}+794 t^{14}-1491 t^{13}-559 t^{12}+3597 t^{11}+18745 t^{10}+38767 t^9+54123 t^8+54123 t^7+38767 t^6+18745 t^5+3597 t^4-559 t^3-1491 t^2+794 t-126\right)}{(t-1)^{11} t^6}\\
&\frac{2 q^5 \left(336 t^{17}-1662 t^{16}+1810 t^{15}+2173 t^{14}+1337 t^{13}-21131 t^{12}-65159 t^{11}-122387 t^{10}-162607 t^9-162607 t^8-122387 t^7-65159 t^6-21131 t^5+1337 t^4+2173 t^3+1810 t^2-1662 t+336\right)}{(t-1)^{11} t^7}\\
&+O(q^6)
\end{split}\]
It agrees with the expansion from \cite{AABN}.

Another interesting consistency check gives comparison of the functions $Z(t,q)$ and $\Psi(t,q)$ at $t=-1$. Literal comparison is not very fruitful because by virtue of  (\ref{E:prodequation2})  $Z(-1,q)=\Psi(-1,q)=0$. 
To get a nonzero result, we used {\it Mathematica} to compute the derivatives  $\partial_t Z(-1,q)$ and $\partial_t \Psi(-1,q)$. Two values  agree by giving 
 \begin{equation}\label{E:zminusone}
 \begin{split}
 &-1024 (q,q)^{22}_{\infty}\partial_t Z(-1,q)=-1024 (q,q)^{22}_{\infty}\partial_t \Psi(-1,q)=1 - 48 q + 1104 q^2 - 16192 q^3 + 170064 q^4 - 1362336 q^5 + \\
 &8662720 q^6 - 44981376 q^7 + 195082320 q^8 +O(q^9).\end{split}
 \end{equation}
The coefficients $a_n$ (up to a sign)  coincide with  the sequence  \href{http://oeis.org/A000156}{A000156} 
. $a_n$, according to the database, is the number of ways of writing $n$ as a sum of $24$ squares. 
This is why it is very plausible that the series (\ref{E:zminusone}) is the expansion of 
\[\left(\sum_{n\in \Z }(-1)^nq^{n^2}\right)^{24}
=(1-q)^{24}(q,q^2)^{24}_{\infty}(q^2,q^2)^{24}_{\infty}(q^3,q^2)^{24}_{\infty}.\]
\section{Functions $\theta_k$}
$\theta_k,k=1,2,3$ are closely related to Rogers-Selberg functions  (see e.g. \cite{Slater},\cite{Andrews}\cite{Hahn}). They satisfy

\[A(q):=\sum_{n\geq0}\frac{q^{2n^2}}{(1-q^2)(1-q^4)\cdots (1-q^{2n})(1+q)(1+q^2)\cdots (1+q^{2n})}=\frac{(q^3,q^7)_{\infty}(q^4,q^7)_{\infty}(q^7,q^7)_{\infty}}{(q^2,q^2)_{\infty}}=\frac{\theta_1}{2(q^2,q^2)_{\infty}},\]
\[B(q):=\sum_{n\geq0}\frac{q^{2n^2+2n}}{(1-q^2)(1-q^4)\cdots (1-q^{2n})(1+q)(1+q^2)\cdots (1+q^{2n})}=\frac{(q^2,q^7)_{\infty}(q^5,q^7)_{\infty}(q^7,q^7)_{\infty}}{(q^2,q^2)_{\infty}}=-\frac{\theta_2}{2(q^2,q^2)_{\infty}},\]
\[C(q):=\sum_{n\geq0}\frac{q^{2n^2+2n}}{(1-q^2)(1-q^4)\cdots (1-q^{2n})(1+q)(1+q^2)\cdots (1+q^{2n+1})}=\frac{(q,q^7)_{\infty}(q^6,q^7)_{\infty}(q^7,q^7)_{\infty}}{(q^2,q^2)_{\infty}}=\frac{\theta_3}{2(q^2,q^2)_{\infty}}.\]
For  a list known identities these functions  obey see \cite{Hahn}.
\section{$Z(t,q,g)$ in the general case}
Some of the above arguments extend to  $Z(t,q,g)$. In particular the divisor of poles of $Z(t,q,g)$ satisfies equations \[1-q^ntv_{\ralpha}(z)=0\] where $v_{\halpha}(z)$ are the weights of the spinor representation. In order to be more explicit recall (see see e.g. \cite{LocMov}) that coordinates $\lambda^{\ralpha}$ on the spinor representation of $\Spin(10)$ can be labelled by elements of the set \[\ralpha\in \setE:= \{(0),(ij),(k)|1\leq i<j\leq 5,1\leq k \leq 5\}\]
Let $\widetilde{\bT^5}$ be the two sheeted cover of the maximal torus $\bT^5\subset \SO(10)$ and $z=(z_1,\dots,z_5)$ be the image of $g$ under projection $\widetilde{\bT^5}\to \bT^5$. The action $\Pi$ of  $g\in \widetilde{\bT^5}$ on $\lambda^{\ralpha}$ is given by the formula
\begin{equation}\label{E:weights}
\begin{split}
&\Pi(g)\lambda^{\ralpha}v_{\ralpha}(z)\lambda^{\ralpha},\\
&\text{{\rm or in more details: }}\\
&\Pi(g)\lambda^{(0)} ={\det}^{-\frac{1}{2}}(z)\lambda^{(0)},\\
&\Pi(g)\lambda^{(ij)}={\det}^{-\frac{1}{2}}(z)z_iz_j\lambda^{(ij)},\\
&\Pi(g)\lambda^{(k)}={\det}^{\frac{1}{2}}(z)z^{-1}_k\lambda^{(k)},\\
&{\det}^{\frac{1}{2}}(z)=\sqrt{z_1\cdots z_5}.
\end{split}
\end{equation}
Introduce the product
\[\theta_{\setE}(t,q ,z):= \prod_{\ralpha\in \setE}\theta(tv_{\ralpha}(z), q)\]
If the conjecture about the structure of the poles of $Z(t,q,z)$ is correct, then 
\[\Xi(t,q,z):=t^6Z(t,q,z) \theta_{\setE}(t,q ,z)\]
 is an analytic function for $|q|<1$ and $t,z\in \C^{\times}\times \widetilde{\bT^5}$. Equations (\ref{E:bgequations1}) (\ref{E:thetasmall}) imply that $\Theta(t,q,z)$ satisfies
 \[\Xi(qt,q,z)=\frac{1}{t^{12}q^{12}}\Xi(t,q,z)\]
That is $\Xi(t,q,z)$ is a section of $\L^{\otimes 12}$.
 Let us fix a basis 
\[\eta_k(t,q)=\sum_{n\in\Z}q^{\frac{12n(n+1)}{2}+kn}t^{12n+k}, k=0,\dots, 11\]
in the space of global sections of $\L^{\otimes 12}$.
Functions $\eta_k$ satisfy \[\eta_k(qt,q)=1/(tq)^{12}\eta_k(t,q),\quad \eta_k(1/t,q)=q^{12-2k}t^{12}\eta_{12-k}(t,q),\quad \eta_{k}(t,q)=q^{k+12}\eta_{k+12}(t,q)\]
The function $\Xi(t,q,z)$ is a linear combination
\[\Xi(t,q,z)=\sum_{k=0}^{11}c_k(q,z)\eta_k(t,q).\]
(\ref{E:prodequation2}) implies that 

\[c_k(q,z)=-q^{12-2k}c_{12-k}(q,z^{-1}), k\neq 0,6,\]
\[c_0(q,z)=-c_{0}(q,z^{-1}),\quad c_6(q,z)=-c_{6}(q,z^{-1})\]
Determination of the coefficients $c_i(q,z)$ is more difficult than in the case $z=1$ and will be postponed for the future publications.

Note that after specialization $g=1$   $\Xi(t,q,1)=\Theta(t,q)\theta(t,q)^5$ and $\theta_{\setE}(t,q ,1)=\theta(t,q)^{16}$ giving as a fraction  the function $Z(t,q)$.

\section{Concluding remarks}
Partition function $Z_{\tspace}$ for $\tspace=\qu$ a smooth affine  quadric  of dimension $n-1$   is known \cite{AA}:
\[Z_{\qu}(t,q)=\frac{1-t^2}{(1-t)^{n}} \frac{(qt^2,q)_{\infty}(qt^{-2},q)_{\infty}}{(qt,q)^n_{\infty}(qt^{-1},q)^n_{\infty}}\]
It satisfies
\begin{equation}\label{E:thetaquad}
\frac{Z_{\qu}(qt,q)}{Z_{\qu}(t,q)}=(-1)^{n}t^{n-4}q^{-1},
\end{equation}
\begin{equation}\label{E:invquad}
 \frac{Z_{\qu}(t^{-1},q)}{Z_{\qu}(t,q)}=-(-t)^{n-2}.
\end{equation}
Functions $Z_{\cone}(t,q),Z_{\qu}(t,q)$ have some common features. 
In both cases $\lim_{t\to1}Z_{\tspace}(t,q)(1-t)^{\dim \tspace}=c_{\tspace}\frac{1}{(q,q)^{\dim \tspace}}$, where $c_{\tspace}$ is some constant. Functions 
$Z_{\tspace}(t,q)$ have poles of multiplicity $\dim\tspace$ at points $\{q^k\}$.

To further extend the  analogy we need to digress. The spaces of polynomial maps $\C\to \tspace$ of degree $N$ is a cone over the space of Drinfeld's quasimaps $QMaps_N(\tspace)$ to projectivization of $\tspace$. The space $QMaps_N(\tspace)$ is not smooth but still has a well defined line bundle of algebraic volume forms $\K$. $\K^{*}=\O(a(\tspace)+N b(\tspace))$ in for some constants $a(\tspace),b(\tspace)$. As usual $\O(n)$ is the power of the tautological line bundle.  Exponents of $t$ in (\ref{E:bgequations1}) and (\ref{E:thetaquad}) coincide with $b(\cone)$ and $b(\qu)$.  Thus functions $\Theta_{\cone}(t,q)$ and $\Theta_{\qu}(t,q)$ are sections of $\L^{\dim\cone-a(\cone)}=\L^{7}$ and $\L^{\dim\qu-a(\qu)}=\L^{3}$ respectively.
Finally $P_\tspace(t)=\lim_{q\to 0}Z_{\tspace}(t,q)$ is the classical Poincar\'{e} series of the algebra of homogeneous functions on $\tspace$. 

It is tempting to say that this is a common features of an elliptic generalization of Poincar\'{e} series of an algebra of functions $\tspace$ that should exist for a  class of conical  varieties whose members are  $\cone$ and $\qu$. This class contains in the class of local conical  Calabi-Yao varieties $\tspace$, whose base $B(\tspace)$  is Fano of sufficiently large index. In this generalization \[Z_{\tspace}(t,q)=\frac{\Theta_{\tspace}(t,q)}{t^{l(\tspace)}\theta(t,q)^{\dim \tspace}}.\]
$l(\tspace)\in \Z^{>0}$, $\Theta_{\tspace}$ is a section of $\L^{\dim\tspace-a(\tspace)}$. The denominator in this formula is similar to the denominator in the Kac formula for the character of an integrable representations of an affine Lie algebra $\hat{\g}$ at a positive level.

It appears that the first nontrivial  Laurent coefficients of $Z_{\qu_n}(t,q)$ for a  quadric at $t=1$ can be expressed through algebraic combinations of $(q,q)_{\infty}, E_2, E_4$ and probably $E_6$ . The  formulas are  similar to the ones that have already appeared in  Section \ref{E:evidences}:
\[\begin{split}
&\lim_{t\to1}Z_{\qu_n}(t,q)(1-t)^{n-1}=2(q,q)_{\infty}^{2-2n}\\
&\lim_{t\to1}\partial_t(Z_{\qu_n}(t,q)(1-t)^{n-1})=(q,q)_{\infty}^{2-2n}\\
&\lim_{t\to1}\partial^2_t(Z_{\qu_n}(t,q)(1-t)^{n-1})=(q,q)_{\infty}^{2-2n}\frac{(n-4)(1-E_2)}{6}\\
&\lim_{t\to1}\partial^3_t(Z_{\qu_n}(t,q)(1-t)^{n-1})=(q,q)_{\infty}^{2-2n}\frac{(4-n)(1-E_2)}{4}\\
&\lim_{t\to1}\partial^4_t(Z_{\qu_n}(t,q)(1-t)^{n-1})=(q,q)_{\infty}^{2-2n}\left(\frac{ (n-4)^2E_2^2}{24} -\frac{(n-4) (n+6)E_2}{12}  +\frac{(n-16)E_4}{60} +\frac{5 n^2+58 n-288}{120}\right)\\
\end{split}\]
Introduce an increasing multiplicative  filtration on the algebra of function generated by $E_2,E_4,E_6$. The $n$-th derivative of $Z_{\qu_n}(t,q)(1-t)^{n-1}$ at $t=1$ is a multiple of  $(q,q)_{\infty}^{-2\dim\qu}$. The factor belongs to $n$ filtration space of the algebra.  This parallels between $Z_{\qu_n}$ and  $Z_{\cone}$ suggests that this also holds  for  a more general $\tspace$.
It raises a  question whether coefficients of the $E_2,E_4,E_6$-monomials in the above formulas   can be computed in terms of the characteristic classes of some bundles on 
the base of the cone $\tspace$.

\end{document}